\documentclass[letterpaper]{article} 
\usepackage[]{aaai2026}  
\usepackage{times}  
\usepackage{helvet}  
\usepackage{courier}  
\usepackage[hyphens]{url}  
\usepackage{graphicx} 
\usepackage{natbib}  
\usepackage{caption} 
\usepackage{algorithm}
\usepackage{algorithmic}
\usepackage{booktabs}
\usepackage{multirow}
\usepackage{tabularx}
\usepackage[table,xcdraw]{xcolor}
\usepackage{caption}
\usepackage{subcaption}
\usepackage{amsmath}
\usepackage{amssymb}
\usepackage{enumitem} 
\usepackage[T1]{fontenc}

\definecolor{prompt_var}{HTML}{8A2BE2}  
\definecolor{prompt_context}{HTML}{556B2F} 

\usepackage{newfloat}
\usepackage{listings}
\DeclareCaptionStyle{ruled}{labelfont=normalfont,labelsep=colon,strut=off} 
\floatstyle{ruled}
\newfloat{listing}{tb}{lst}{}
\floatname{listing}{Listing}
\definecolor{hlgreen}{HTML}{B2D5CB}
\definecolor{hlblue}{HTML}{ADD8E6}
\definecolor{bggrey}{HTML}{5E5D65}
\definecolor{bgpink}{HTML}{CEAEB9}
\definecolor{bgblue}{HTML}{8D91AA}

\nocopyright

\title{The Imperfect Learner: Incorporating Developmental Trajectories \\in Memory-based Student Simulation}

\author{
    Zhengyuan Liu\textsuperscript{\rm 1,*},
    Stella Xin Yin\textsuperscript{\rm 3,*},
    Bryan Chen Zhengyu Tan\textsuperscript{\rm 2},\\
    Roy Ka-Wei Lee\textsuperscript{\rm 2},
    Guimei Liu\textsuperscript{\rm 1},
    Dion Hoe-Lian Goh\textsuperscript{\rm 3},
    Wenya Wang\textsuperscript{\rm 3},
    Nancy F. Chen\textsuperscript{\rm 1}
}
\affiliations{
    \textsuperscript{\rm 1}Institute for Infocomm Research, A*STAR, Singapore\\
    \textsuperscript{\rm 2}Singapore University of Technology and Design\\
    \textsuperscript{\rm 3}Nanyang Technological University, Singapore\\
    {liu\_zhengyuan@a-star.edu.sg, nancy\_chen@a-star.edu.sg}
}

\begin{document}

\maketitle

\begin{abstract}
User simulation is important for developing and evaluating human-centered AI, yet current student simulation in educational applications has significant limitations. Existing approaches focus on single learning experiences and do not account for students' gradual knowledge construction and evolving skill sets. Moreover, large language models are optimized to produce direct and accurate responses, making it challenging to represent the incomplete understanding and developmental constraints that characterize real learners.
In this paper, we introduce a novel framework for memory-based student simulation that incorporates developmental trajectories through a hierarchical memory mechanism with structured knowledge representation. The framework also integrates metacognitive processes and personality traits to enrich the individual learner profiling, through dynamical consolidation of both cognitive development and personal learning characteristics.
In practice, we implement a curriculum-aligned simulator grounded on the Next Generation Science Standards. Experimental results show that our approach can effectively reflect the gradual nature of knowledge development and the characteristic difficulties students face, providing a more accurate representation of learning processes.
\end{abstract}

\section{Introduction}
The rapid advancement of generative AI has raised much interest in research and application-driven works across educational contexts \cite{kasneci2023chatEDU,law2024application}. Built on multi-modal and multilingual foundation models \cite{achiam2023gpt}, the intelligent systems aim to transform how students engage with learning materials, receive personalized feedback, and develop understanding across various subjects \cite{zhang2024systematic,chiu2024impact}. However, there is an urgent need for efficient and scalable methods to evaluate and improve these automated systems.
Testing AI-based solutions with real students faces several challenges: recruitment is time-consuming and costly, ethical considerations limit research scope, early-stage systems may pose potential risks to learners, and requiring diverse participants makes comprehensive evaluation difficult. User simulation offers an alternative approach, enabling researchers to scale up early participant numbers across varied learning scenarios without the logistical and ethical constraints of working with actual students \cite{lu2024generative, liu-etal-2024-personality, liu2025leveraging}.

\begin{figure}[t!]
\centering
\includegraphics[width=0.92\linewidth]{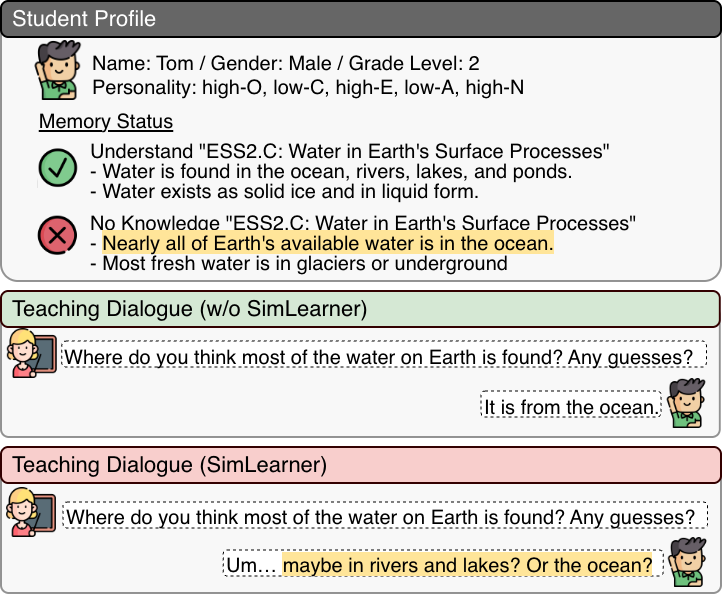}
\caption{One example of memory-based student simulation: When the knowledge setting is on basic water concepts and the student is asked a harder question, SimLearner demonstrates uncertainty and guessing behavior (as highlighted), while the base method directly provides the correct answer.}
\label{Figure_Intro}
\vspace{-0.4cm}
\end{figure}

While existing work in role-playing with large language models has shown success in simulating human personas and behaviors \cite{zhang2024exploring}, educational contexts present unique challenges that distinguish student simulation from general role-playing tasks. Unlike adult professionals or personas with established knowledge bases, it requires capturing incomplete understanding, developmental constraints, and the gradual process of knowledge construction that characterizes authentic learning experiences.
For instance, elementary school students demonstrate concrete thinking, express misconceptions, show variable attention spans, and express emotional responses to learning challenges \cite{Piaget1971}. Presenting these characteristics requires fine-grained and dynamic state tracking rather than simplified versions of response generation \cite{Flavell1999}, and challenges come from three aspects:
(1) Large language models (LLMs), despite their impressive capabilities in language understanding and generation \cite{safdari2023personality}, face an inherent challenge in role-playing imperfect learners (as the example shown in Figure \ref{Figure_Intro}). LLMs are optimized to produce direct and accurate responses that reflect concrete reasoning \cite{tseng2024two}. Incorporating the incomplete understanding, developmental limitations, and forgetting patterns of elementary learners requires approaches that go beyond injecting random mistakes to capture how children actually construct and express their understanding of new concepts \cite{Gopnik1999}.
(2) Existing student simulations typically simulate single learning interactions rather than the continuous, gradual knowledge construction \cite{liu-etal-2024-personality}. Students do not simply progress from incorrect to correct answers in discrete steps. Instead, they build understanding incrementally, exhibit evolving misconceptions, and demonstrate learning patterns that change over extended periods \cite{Piaget1952,carey2000science}.
(3) Most student simulation methods lack the individual variation that defines real classrooms. Students differ not only in their knowledge states but also in their personality traits \cite{Costa1999,John1999}, learning preferences \cite{dunn1984learning}, and metacognitive skills \cite{fisher1998thinking}. These individual differences shape how students engage with content, respond to challenges, and develop understanding over time \cite{Busato1998,Komarraju2011}.

In this work, we propose a novel framework for student simulation that addresses these limitations. To incorporate developmental trajectories, we build a hierarchical memory mechanism with structured knowledge representation that captures the gradual, linked nature of knowledge construction. By leveraging fine-grained memory retention, updating, forgetting, and tracing, this architecture enables models to demonstrate grade-appropriate capabilities and learning preferences.
Moreover, we enrich individual learner profiles with personality traits \cite{Busato1998} and metacognitive skills \cite{fisher1998thinking} to reflect cognitive development and noncognitive learning characteristics.

We demonstrate the framework's effectiveness by implementing a student simulator \textbf{SimLearner} in elementary school science learning grounded in the Next Generation Science Standards (NGSS) \cite{bybee2014ngss,liu-etal-2025-cogent}. Experimental results demonstrated that our approach aligns better with the curriculum standards of grade-adapted learning trajectories, and captures the characteristic difficulties that elementary students encounter when engaging with science concepts.

Our work moves beyond simple context engineering to create a sophisticated simulation system by integrating learning science, pedagogical principles, and psychology theories upon LLMs capabilities. Our framework provides researchers and educators with a tool for evaluating educational interventions with diverse student populations and learning scenarios. By building on the imperfect, gradual, and highly individual nature of learners, the memory-based approach opens new possibilities for understanding and supporting technology integration and application in educational contexts.

\section{Related Work}
\subsection{Model-based User Simulation}

Large language models have shown strong capabilities in simulating human behaviors across diverse domains \cite{zhang2024exploring}. Recent work has explored social interactions \cite{zhou-etal-2024-real,tan-etal-2025-persuasion}, decision-making \cite{yang2024llm}, and mobility patterns \cite{jiawei2024large}. These simulations have shown success in replicating aggregate human behaviors and decision patterns in controlled settings \cite{dorner2023personality}.
However, simulating the specific characteristics of developing learners presents unique challenges. Unlike adult decision-making or social behaviors, student learning involves ongoing cognitive development \cite{Piaget1971}, incomplete knowledge states, and developmental constraints that fundamentally shape how information is processed and expressed \cite{klahr2022cognitive}. Existing methods focus primarily on mature reasoning patterns and may struggle to capture the authentic thinking processes of elementary learners \cite{jiang2023personallm}.

\begin{figure*}[t!]
\centering
\includegraphics[width=0.99\linewidth]{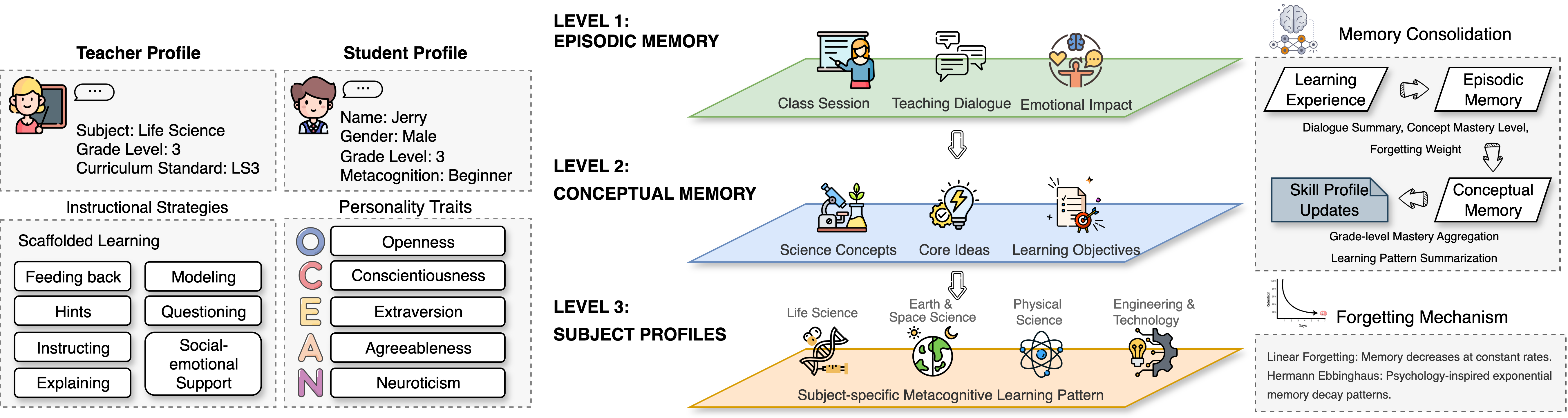}
\caption{Overview of student simulation architecture built on hierarchical memory mechanism, structured knowledge representation, and multi-dimensional profiling (cognitive, noncognitive, metacognitive aspects).}
\label{Figure_Framework}
\end{figure*}

\begin{figure*}[t!]
\centering
\includegraphics[width=0.82\linewidth]{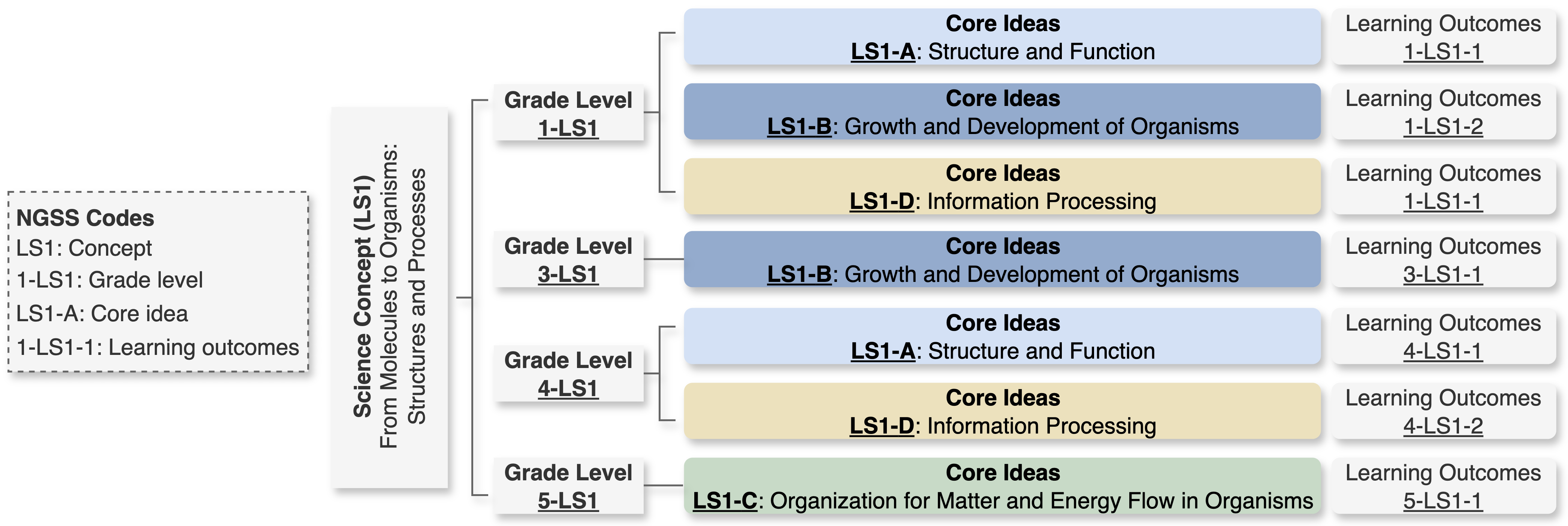}
\caption{The structured knowledge representation grounded on the NGSS curriculum standard.}
\label{Figure_NGSS}
\end{figure*}

\subsection{Student Simulation in Education}
Student simulation has focused primarily on training teaching assistants \cite{liu2024scaffolding}, modeling student responses to assessment items \cite{lu2024generative, liu2025leveraging}, and creating diverse personality-driven agents for educational interactions \cite{liu-etal-2024-personality, gur2018user,asri2016sequence,lin-etal-2022-gentus}. These approaches have made important contributions to understanding how AI can support educational evaluation and teacher training \cite{kreyssig-etal-2018-neural,cheng-etal-2022-multiwoz}.
Existing educational simulations typically model students at single time points rather than incorporating learning trajectories over time \cite{liu-etal-2022-generative,wang2024indepth}. They focus on generating plausible responses to specific educational content without capturing how students' understanding develops through sustained engagement with concepts. This limitation makes it difficult to use these simulations for understanding learning processes or evaluating interventions that depend on developmental change.

\subsection{Memory Architectures in AI Systems}
Recent advances in memory architectures have explored various approaches to storing and retrieving information over extended interactions \cite{zhang2024survey, wu2024autogen}. These include consolidation mechanisms that transform short-term experiences into persistent representations \cite{balavzevic2024memory}, hierarchical memory structures that organize information at different levels of abstraction \cite{lampinen2021towards, sun2025hierarchical}, and graph-based representations that capture relationships between concepts \cite{emmons2020sparse}, where summarization and context retrieval are basic components \cite{liu-2019-topic,liu-etal-2021-coreference,long-etal-2025-reinforcing}.
While these memory architectures provide important foundations for dynamic and persistent context retrieval and processing, they are typically designed to maximize information retention and retrieval accuracy \cite{maclellan2022domain, zhong2024memorybank, du2025rethinking}. For student simulation, we need memory-based designs that capture the partial, evolving, and sometimes incorrect understanding that characterizes real learning patterns.

\section{Methodology}
Our framework addresses the core challenges of student simulation through three integrated components: the structured knowledge representation that guides developmental constraints, the hierarchical memory mechanism that captures evolving states, and the multi-dimensional profiling that reflects noncognitive and metacognitive factors.

\subsection{Structured Knowledge Representation}
In educational scenarios, structured knowledge representation follows a natural hierarchy from broad subjects to specific points \cite{holthuis2018new}, and provides a scaffolded way to enhance student's learning experience, processes and reach targeted learning objectives \cite{anderson2018designing}.
It is commonly organized in multiple levels: \textbf{subjects} represent domain areas (e.g., mathematics, science), \textbf{concepts} capture fundamental ideas within domains (e.g., fractions, photosynthesis), \textbf{core ideas} organize related concepts into coherent themes (e.g., number sense, energy transfer), and \textbf{learning outcomes} specify measurable skills and understanding targets. This hierarchical organization mirrors curriculum design principles and enables precise tracking of student progress across different granularities of knowledge. Each level maintains prerequisite relationships and developmental constraints, ensuring that simulated learning follows pedagogically sound progressions from foundational concepts to advanced applications. For example, a concept like ``forces and motion'' appears in simplified form in early elementary grades focusing on pushes and pulls, then develops into more sophisticated understanding of balanced and unbalanced forces in later grades. This progression allows us to represent partial understanding appropriately, where students might grasp age-appropriate aspects of a concept while lacking understanding from more advanced dimensions.

\subsection{Hierarchical Memory Mechanism}
In learning science theory, students' learning and knowledge acquisition is a progressive process. They gradually construct understanding over multiple learning interactions rather than treating each learning moment as an isolated event \cite{LI2025104206}. To mirror the natural progression from single learning experiences to consolidated knowledge and high-level skills, we propose a three-level hierarchical memory architecture, which can distinguish between working memory and long-term memory processes, facilitating selective knowledge retention, forgetting, updating, and tracing \cite{liu2021hierarchical,miao2025graph}.

As shown in Figure \ref{Figure_Framework}, the memory part operates across three hierarchical levels. The first level maintains \textbf{episodic memory} as temporal sequences of learning experiences, with each unit containing summarized session, learning content, and mastery metadata. A memory strength parameter implements natural forgetting while preserving recent experiences for retrieval \cite{du2025rethinking}.
The second layer represents evolving \textbf{conceptual memory}. Each concept node maintains dynamic understanding descriptions and mastery levels from beginner to fully understanding, with evidence sources linking to contributing episodic memories and grade-level curriculum alignment.
The third level captures \textbf{subject profiles}. These abstractions encompass domain-specific concepts while maintaining learning level indicators and pattern descriptions derived from observed behaviors across recent episodes. Metacognitive skills function as regulatory mechanisms enabling students to identify knowledge gaps and select appropriate strategies \cite{fisher1998thinking}.
The architecture supports both bottom-up consolidation, where episodic experiences inform concept development, and top-down influence, where existing subject-level mastery and metacognitive skills shape how existing experiences are leveraged.

\subsection{Multi-dimensional Learner Profiling}
We also enrich learner profiles by integrating noncognitive and metacognitive characteristics.

\subsubsection{Noncognitive Characteristics}
Personality traits significantly influence learning and memory processes \cite{eysenck1981learning}. Previous studies show that higher Openness, Extraversion, and Agreeableness correlate with better memory performance, while higher Conscientiousness and Neuroticism correlate with poorer retention \cite{WARIS201826}. Building on Big Five theory, personality traits are incorporated for educational contexts: High openness manifests as creativity in answers and curiosity about learning, while low conscientiousness appears as disorganized responses and easy distraction from tasks.

\subsubsection{Metacognitive Characteristics}
Metacognition consists of metacognitive knowledge (understanding one's cognitive capabilities) and metacognitive regulation (monitoring and evaluating cognitive processes) \cite{lai2011metacognition}. This enhances learning by providing executive control over encoding, storage, and retrieval processes \cite{mazancieux2023towards}.
We set up the key metacognitive skills in three learning phases: forethought and planning, performance monitoring, and self-reflection. Since metacognitive skills evolve along with the learning progress, we summarize the episodes to identify metacognitive behaviors like goal setting and strategy planning, then maintain such characteristic patterns for different subject domains.

\section{Student Simulation for Science Learning}
In this work, we demonstrate the effectiveness of our framework by implementing an elementary school student simulator \textbf{SimLearner} for science learning grounded in the Next Generation Science Standards (NGSS) curriculum.

\subsection{Structured Knowledge Representation}
We convert the Next Generation Science Standards (NGSS) \cite{bybee2014ngss,liu-etal-2025-cogent}\footnote{https://www.nextgenscience.org/} to a curriculum-aligned knowledge structure, which defines a three-level decomposition that maps naturally to the hierarchical memory mechanism, as shown in Figure \ref{Figure_Framework} and Figure \ref{Figure_NGSS}.

\subsubsection{NGSS Curriculum Hierarchy}
We convert the NGSS curriculum standard $\mathcal{K}_{\text{NGSS}}$ to a three-level structure:
$\mathcal{K}_{\text{NGSS}} = \{\mathcal{S}, \mathcal{C}, \mathcal{L}\}$,
where $\mathcal{S}$ represents science subjects, $\mathcal{C}$ denotes concepts, and $\mathcal{L}$ contains learning outcomes that integrate both core ideas and performance expectations.

Science Subjects ($\mathcal{S}$): The highest abstraction level encompasses four primary domains for elementary education (grades 1-5):
$\mathcal{S} = \{\text{LS}, \text{PS}, \text{ESS}, \text{ETS}\}$, corresponding to Life Science, Physical Science, Earth and Space Science, and Engineering Technology Systems, respectively. Each subject $s \in \mathcal{S}$ provides the broad disciplinary context within which specific concepts are situated.

Concepts ($\mathcal{C}$): Each subject decomposes into specific concepts $c_i \in \mathcal{C}$. For each concept element
$c_i = \{\text{id}^{(i)}, \text{desc}^{(i)}, \text{subject}^{(i)}\}$, $\text{id}^{(i)}$ is a unique identifier (e.g., ``LS1''), $\text{desc}^{(i)}$ is the natural language description, and $\text{subject}^{(i)} \in \mathcal{S}$ indicates the parent science subject. Our implementation includes 29 distinct concepts across elementary grades.

Learning Units ($\mathcal{L}$): The most specific level combines core ideas with measurable performance expectations $l_j \in \mathcal{L}$ that students should achieve at each grade level. Each element is
$l_j = \{\text{id}^{(j)}, \text{core\_idea}^{(j)}, \text{outcome}^{(j)}, \text{grade}^{(j)},\}$.
$\text{id}^{(j)}$ is the unique NGSS identifier, $\text{core\_idea}^{(j)}$ specifies the knowledge points, $\text{outcome}^{(j)}$ defines the performance expectation, $\text{grade}^{(j)} \in \{1,2,3,4,5\}$ indicates the corresponding grade level.

\subsection{Hierarchical Memory Mechanism}
This component consists of episodic memory ($M_e$), conceptual memory ($M_c$), and metacognitive skill profiles ($M_s$), each operating at different temporal and abstraction scales.

\subsubsection{Episodic Memory ($M_e$)}
The episodic memory level stores individual learning experiences as discrete memory units $m_e^{(i)}$. Each episodic memory unit is formed as:
$$m_e^{(i)} = \{t^{(i)}, \text{content}^{(i)}, \text{summary}_e^{(i)}, \mathcal{I}^{(i)}, \mathcal{C}^{(i)}, \mathcal{M}^{(i)}, \sigma^{(i)}\}$$
where $t^{(i)}$ represents the timestamp, $\text{content}^{(i)}$ contains the raw dialogue, $\text{summary}_e^{(i)}$ provides a natural language summary, $\mathcal{C}^{(i)}$ lists relevant NGSS concept identifiers, $\mathcal{M}^{(i)}$ maps concept mastery level, and $\sigma^{(i)} \in [0,1]$ represents memory strength for forgetting dynamics.

Consolidation from dialogue to episodic memory employs an LLM-based summarizer that extracts key insights:
$$\text{summary}_e^{(i)}, \mathcal{I}^{(i)}, \mathcal{C}^{(i)}, \mathcal{M}^{(i)} = \text{LLM}_{\text{consolidate}}(\text{dialogue}^{(i)}, \mathcal{N})$$

where $\mathcal{N}$ represents the NGSS standards taxonomy and $\text{LLM}_{\text{consolidate}}$ performs structured extraction of learning insights (see details in Supplementary Appendix).

\begin{figure}[t!]
\centering
\includegraphics[width=0.8\linewidth]{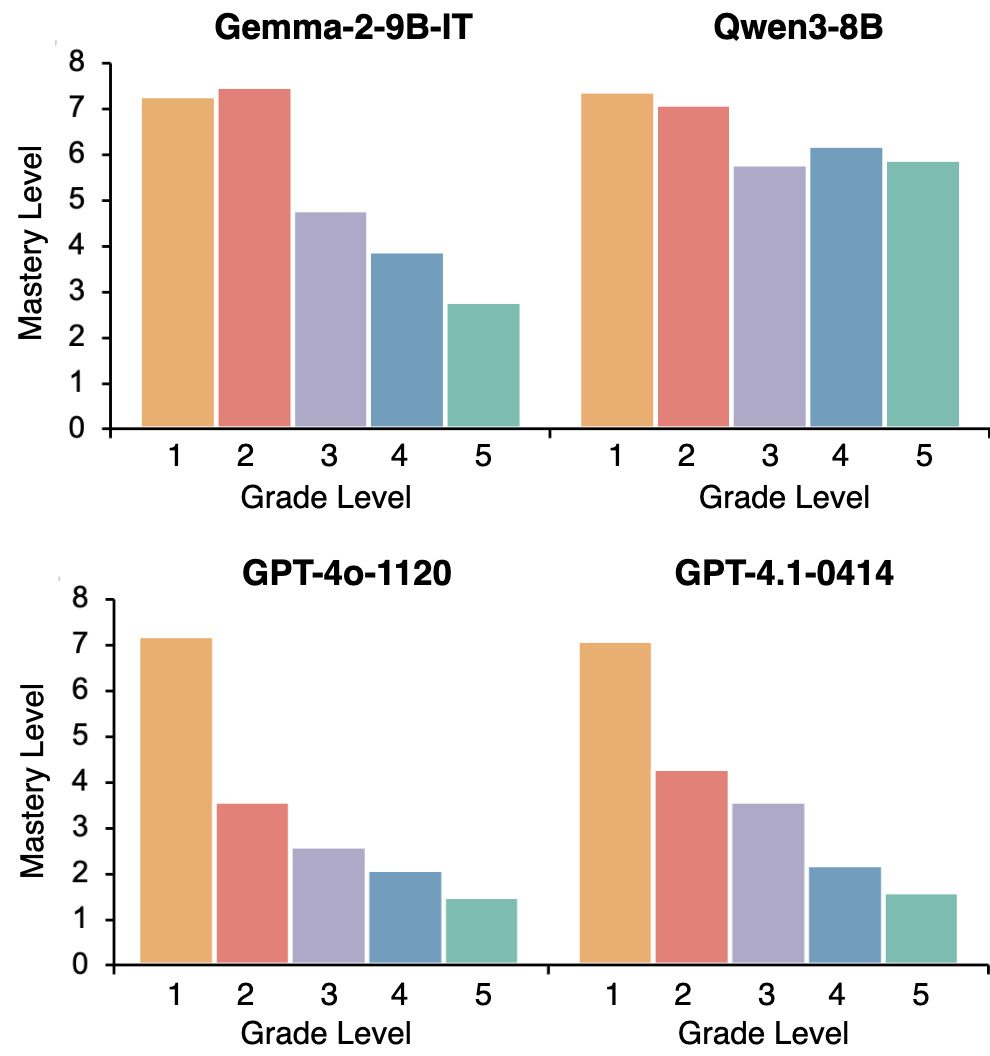}
\caption{Results of knowledge mastery simulation as a grade-1 student. Lower scores in grades 2-5 indicate better alignment with natural developmental levels.}
\label{Figure_QA}
\end{figure}

\subsubsection{Conceptual Memory ($M_c$)}
This level maintains understanding states for individual NGSS concepts through concept nodes $m_c^{(j)}$. Each concept node presents the student's current understanding and mastery progression:
$$m_c^{(j)} = \{id^{(j)}, desc^{(j)}, \mu^{(j)}, \mathcal{E}^{(j)}, g^{(j)}\}$$

where $id^{(j)}$ denotes NGSS concept id, $desc^{(j)}$ is concept description, $\mu^{(j)} \in [0,1]$ quantifies mastery score, $\mathcal{E}^{(j)}$ linked to related episodes, and $g^{(j)}$ is the grade level.

Consolidation from episodic to conceptual memory implements both mastery aggregation and forgetting mechanisms. We implement a naive forgetting mechanism (the Ebbinghaus curve can also be applied).
For each concept $j$ referenced in episode $i$, the mastery update follows:
$$\mu^{(j)}_{t+1} = \alpha \cdot \mu^{(j)}_t + \beta \cdot w^{(i,j)}$$
where $\alpha$ represents the forgetting decay factor (we set it at 0.95 per episode), $\beta$ controls the learning rate (here we set it at 0.25 according to the averaged episode number of each subject per grade), and $w^{(i,j)}$ weights the demonstrated mastery from episode $i$ for concept $j$. This forgetting mechanism gets updated at every episode consolidation process, with the decay rate influenced by grade-level appropriateness, demonstrated mastery level, and exposure frequency. This creates realistic patterns where inappropriate or unused knowledge fades over time while actively engaged concepts strengthen through repeated exposure.

\subsubsection{Metacognitive Skill Profiling ($M_s$)}
The highest level captures learning patterns and metacognitive strategies through skill profiles $m_s^{(k)}$ for different subject domains:
$$m_s^{(k)} = \{subj^{(k)}, level^{(k)}, pattern^{(k)}, g^{(k)}\}$$

where $subj^{(k)}$ is the subject domain, $level^{(k)} \in \{\text{beginner, developing, expert}\}$ represents overall competency, $pattern^{(k)}$ contains a natural language description of learning patterns, and $g^{(k)}$ denotes the current grade level.

The consolidation process evaluates three phases of self-regulated learning. For $n$ recent episodes from $m_e^{(i)}$ to $m_e^{(i+n)}$, an LLM-based summarizer extracts learning patterns from the dialogues and mastery levels by categorizing meta-cognitive skills of forethought, performance, and reflection (see details in Supplementary Appendix).

\begin{figure}[t!]
\centering
\includegraphics[width=0.8\linewidth]{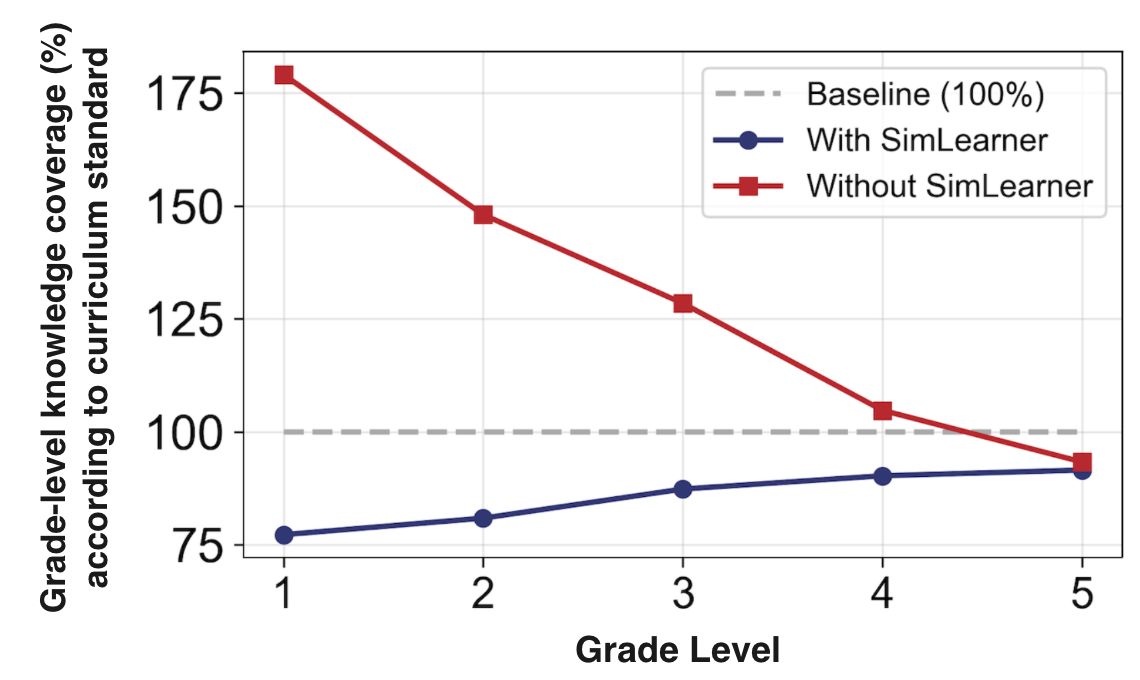}
\caption{Comparison of grade-level knowledge coverage according to the curriculum standard from grade 1 to 5. Compared with the base method (red line), SimLearner shows a much smaller gap with the expected coverage.}
\label{Figure_Grade_Coverage}
\end{figure}

\subsubsection{Personality-aware Profiling}
We adopt the Big Five for Tutoring Conversations (BF-TC) personality scheme proposed in \cite{liu-etal-2024-personality}. The controllable personality $\mathcal{P}_{\text{BF-TC}}$ consists of five traits: openness, conscientiousness, extraversion, agreeableness, and neuroticism. The personality context $\phi(\mathcal{P}_{\text{BF-TC}})$ maps trait configurations (e.g., \textit{low}, \textit{high}) to natural language descriptions.

\subsection{Dynamic Context for Student Simulation}
The full context for the student's response generation includes personality $\phi(\mathcal{P}_{\text{BF-TC})})$, memory context $\psi(M_e, M_c, M_s)$, dialogue history $h_{\text{dialogue}}$ and developmental constraints $\theta_{\text{dev}}$:
$$r = \text{LLM}_{\text{respond}}(\phi(\mathcal{P}_{\text{BF-TC}}), \psi(M_e, M_c, M_s), h_{\text{dialogue}}, \theta_{\text{dev}})$$

The $\psi(M_e, M_c, M_s)$ dynamically adjusts knowledge pieces (e.g., $K_{\text{learned}}$, $K_{\text{unknown}}$) based on the student's current grade level described in $\theta_{\text{dev}}$ and dialogue history $h_{\text{dialogue}}$.
Here $K_{\text{learned}}$ represents learned knowledge related to the current session, $K_{\text{unknown}}$ denotes inaccessible advanced concepts.

In particular, concepts above the student's current grade level or marked with low mastery scores are indicated to prevent unrealistic advanced responses. For instance, a 2nd-grade student won't suddenly discuss complex topics like ``energy flow in organisms'' (an NGSS grade 5 concept) that they shouldn't know yet. By explicitly adding ``unknown concepts'' in the context, the simulator behaves more like a real child who only knows what they've actually learned at their developmental stage.

\begin{figure*}[t!]
\centering
\includegraphics[width=1.0\linewidth]{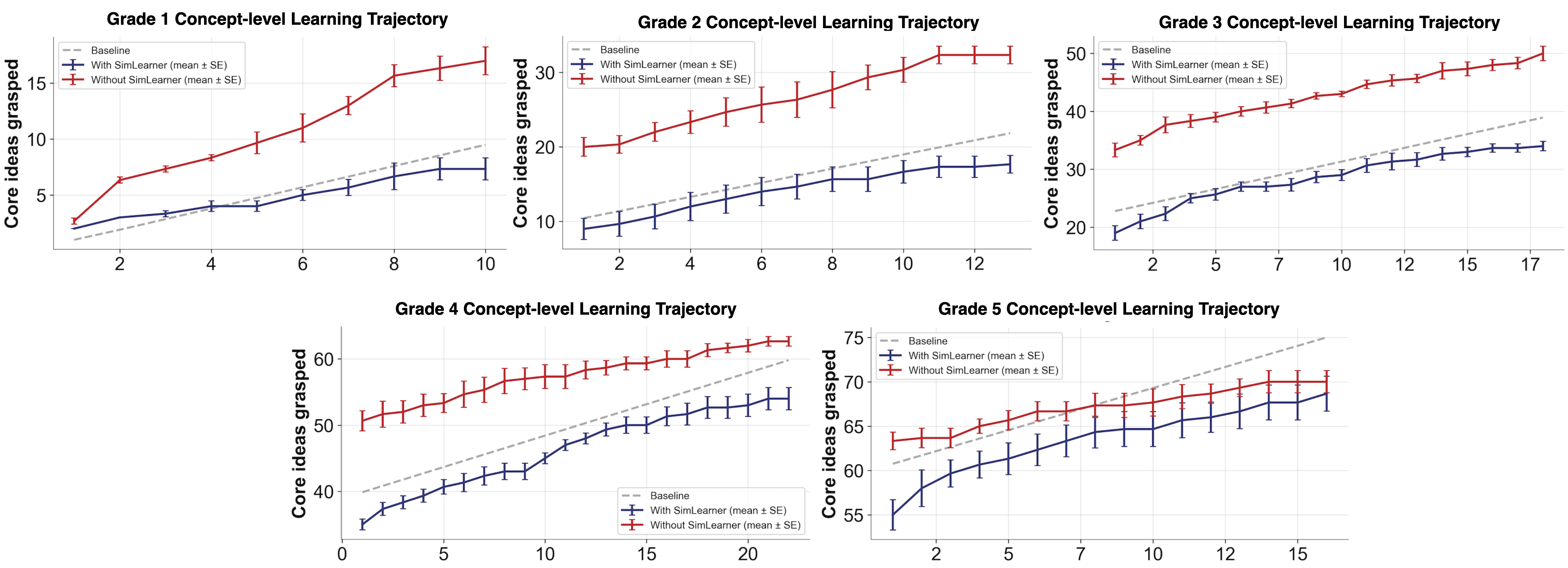}
\caption{Simulated concept-level learning trajectories across elementary grades 1-5. Grey dashed lines represent curriculum standard baselines. Blue lines show students with SimLearner and Red lines show students without SimLearner. Error bars indicate variation across three representative student profiles in each condition.}
\label{Figure_learning_grade}
\end{figure*}

\begin{figure*}[t!]
\centering
\includegraphics[width=1.0\linewidth]{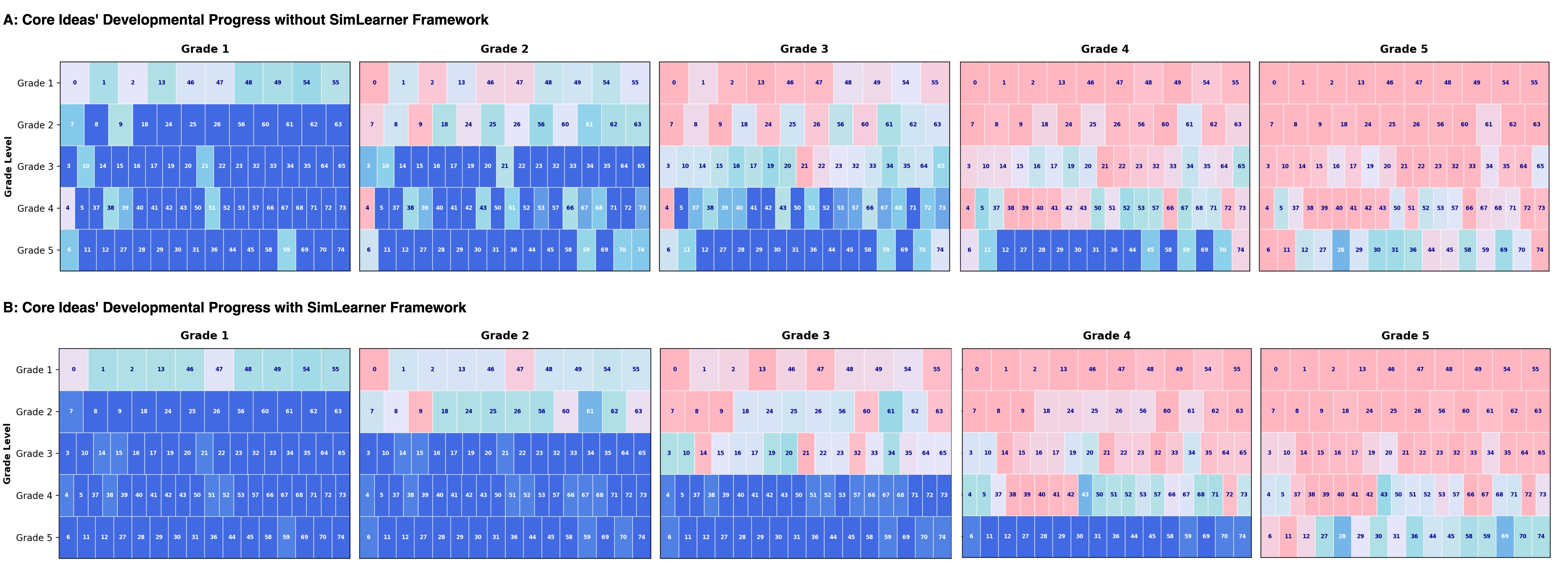}
\caption{Visualization and comparison of NGSS core ideas developmental progress between A) without SimLearner and B) with SimLearner from grade 1 to grade 5. The numbered cells represent core idea IDs specified in NGSS and distributed at each grade level. Blue coloring indicates unfamiliarity with core ideas, while pink coloring indicates familiarity.}
\label{Figure_Heatmap}
\vspace{0.1cm}
\end{figure*}

\section{Experimental Results and Analysis}
We conducted comprehensive assessments across multiple dimensions to validate SimLearner's effectiveness in elementary science learning. In our experiments, the teacher agent selects grade-appropriate science core ideas from the NGSS curriculum, and conducts dialogic tutoring through questioning, explanations, and feedback until the student demonstrates understanding \cite{liu2024scaffolding}. Both agents are tested across grades 1-5, with the student's knowledge and responses adjusted to match developmental expectations for each grade level.
Following previous work \cite{tseng2024two,liu2025leveraging}, we selected and tested four models as the simulator (Gemma-2-9B, Qwen3-8B, GPT-4o, GPT-4.1), and used Claude-Sonnet-3.7 for LLM-as-a-judge. Implementation details are shown in Supplementary Appendix.

\subsection{Knowledge Mastery Consistency}
In real-world educational settings, students demonstrate understanding patterns that align with curriculum-required grade-level learning outcomes, with students of varying abilities showing different levels of comprehension and misconceptions in science classes. To simulate this scenario, we examined students' knowledge mastery levels through question and answer sessions across different grade levels.
We initialized the student agent at grade 1 and the teacher agent was to ask questions from NGSS curriculum items spanning grades 1-5. Here we hypothesized that a ``reasonable'' simulation would demonstrate grade-appropriate performance patterns, answering grade 1 questions correctly and confidently while showing limited knowledge or guessing for questions beyond grade 1. We evaluated the answers on a 1-10 scale to represent knowledge mastery levels, then compared scores across different grade-level question sets. The results revealed significant differences between model types.

As shown in Figure \ref{Figure_QA}, the stronger models (e.g., GPT-4o, GPT-4.1) adhere to grade-aligned capability better. The simulated students answer grade 1 questions confidently with mastery scores around 7, then decrease for higher grade levels, which reflects realistic knowledge acquisition that align with our assumption. In contrast, the two open-source models (Gemma-2-9B, Qwen3-8B) show less consistent grade-appropriate performance. Particularly, Qwen3-8B did not follow the instruction for ``unknown'' knowledge pieces effectively in this case (we speculate that it may require task-specific instruction or that the prior for producing perfect answers is too strong).

\begin{figure}[t!]
\centering
\includegraphics[width=0.75\linewidth]{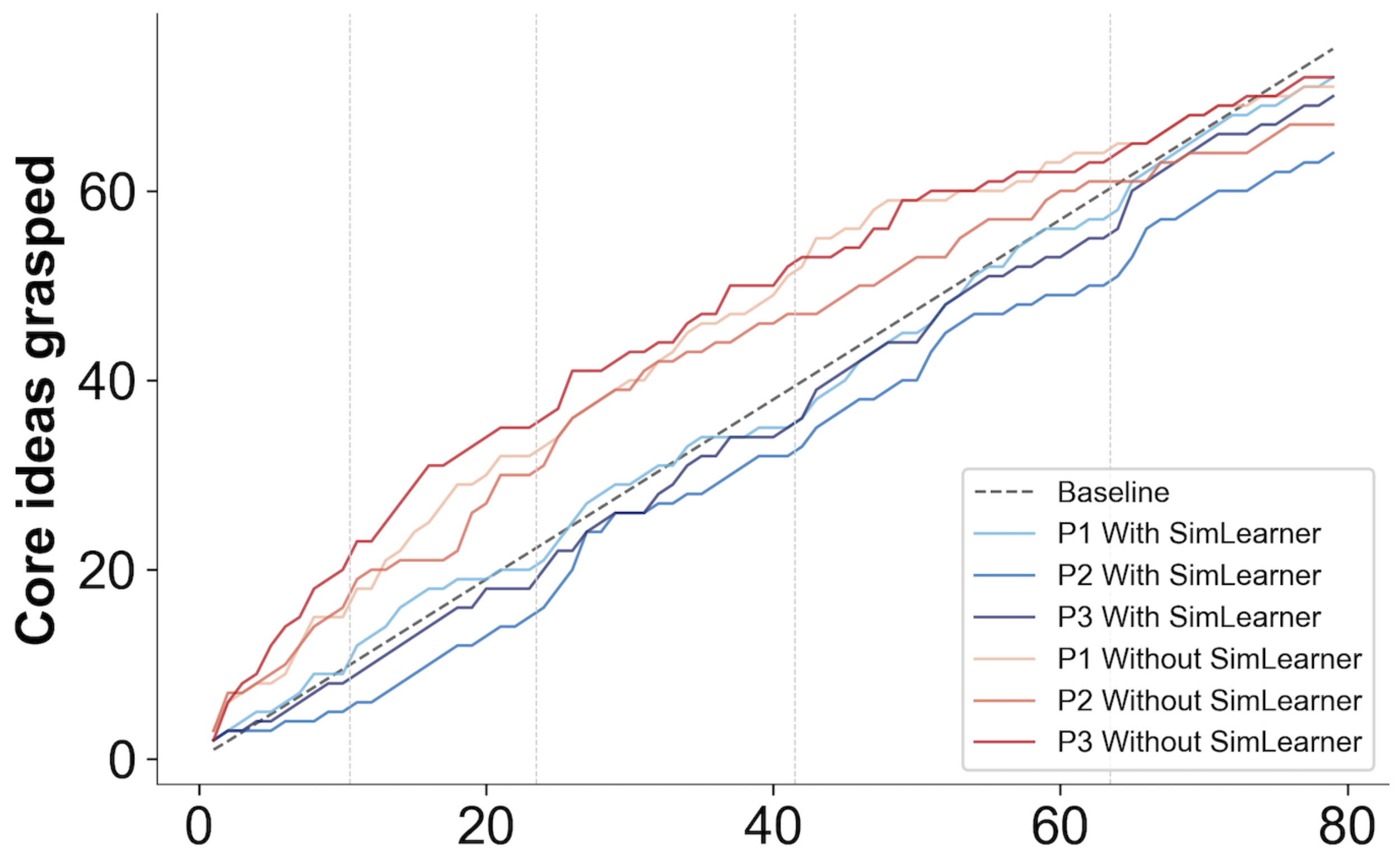}
\caption{Learning trajectories of three student profiles with distinct personality traits across grades 1-5.}
\label{Figure_Personality_Influence}
\vspace{-0.3cm}
\end{figure}

To further validate the effectiveness of LLMs on simulating knowledge boundaries \cite{yao2024large}, we assessed the \textbf{unlearning instruction following} capability by testing whether models could simulate ignorance of specific `unknown' concepts when faced with indirect and adversarial ``trap'' questions. 
Here, we utilized concepts from the NGSS \cite{bybee2014ngss} curriculum as the knowledge source. In each time, we randomly sampled 8 distinct concepts, designating one as the `unknown' target and the remaining as `learned' to simulate a partial knowledge state. We used a teacher agent to generate 10 diverse probe questions related to the `unknown' concept. These questions included direct definition requests, indirect scenarios, and ``trap'' questions designed to tempt the student into revealing the knowledge. The student simulator asked these questions following the developmental constraint. We then evaluated the responses, calculating the \textit{Leakage Ratio}, the proportion of responses where the student failed to demonstrate ignorance and instead provided correct answers or explanations.

Evaluation results reveal significant variance in this particular `unlearning' instruction across different models \cite{liu2025rethinking}, where we tested 100 samples on each LLM. Qwen3-8B exhibited a high leakage ratio of 0.51, indicating that it failed to suppress its intrinsic knowledge when we add the `unknown' constraint. Gemma-2-9B demonstrated improved adherence with a leakage ratio of 0.23. In contrast, GPT variants achieved the robust performance with a minimal leakage ratio around 0.11. This finding suggests that while some models struggle to inhibit their prior knowledge, it is still feasible for such simulation using stronger models that can be conditioned on the cognitive gaps characteristic of developing learners.
These two evaluations become a prerequisite for determining appropriate models for the memory-based simulation.

\subsection{Grade-level Developmental Progress}
We then analyzed how the simulated students with and without SimLearner develop conceptual understanding at each grade level. Following previous work \cite{Komarraju2011}, we defined three distinct student profiles as our experiment examples. \textbf{P1 student} (Gender = Male, Personality = high-O, low-C, high-E, high-A, low-N), \textbf{P2 student }(Gender = Female, low-O, high-C, low-E, low-A, high-N), and \textbf{P3 student} (Gender = Male, high-O, low-C, high-E, low-A, high-N). All students were initially configured with beginner-level metacognitive skills. We conducted experiments where simulated students learned all concepts and core ideas specified in NGSS standards from grades 1 through 5, then observed their learning trajectories under two conditions.

Figure \ref{Figure_learning_grade} presents the results of these learning trajectory comparisons. Here we present the results simulated using GPT-4o. Students with SimLearner (blue) consistently performed within curriculum standards (77.2\% to 91.6\%). Students without SimLearner (red) performed far above grade-level expectations at 179.1\% in grade 1, although gradually converged toward curriculum standards due to the improved difficulty level of concepts. These figures explain how SimLearner maintains developmentally appropriate learning pacing compared to unstructured learning that accelerates students beyond grade-level standards. We further show the knowledge coverage in each grade learning of students with and without the memory-based method (see Figure \ref{Figure_Grade_Coverage}), which highlights the effectiveness of the SimLearner framework in controlling students' cognitive development to align with curriculum requirements.

\begin{table}[t!]
\centering
\small
\begin{tabular}{l|p{0.5cm}<{\centering} p{0.5cm}<{\centering} p{0.5cm}<{\centering} p{0.5cm}<{\centering} p{0.5cm}<{\centering} p{0.5cm}<{\centering}}
\hline
Simulator Model & \multicolumn{3}{c}{Base Method} & \multicolumn{3}{c}{SimLearner} \\
\cline{2-4} \cline{5-7}
 & R & P & F1 & R & P & F1 \\
\hline
Gemma-2-9B-IT & 0.67 & 0.67 & 0.67 & 0.66 & 0.64 & 0.65 \\
Qwen3-8B & 0.65 & 0.64 & 0.64 & 0.63 & 0.62 & 0.62 \\
GPT-4o-1120 & 0.75 & 0.74 & 0.73 & 0.72 & 0.72 & 0.71 \\
GPT-4.1-0414 & 0.74 & 0.74 & 0.73 & 0.71 & 0.70 & 0.70 \\
\hline
\end{tabular}
\caption{Evaluation results on personality simulation. R, P and F1 denotes Recall, Precision, and F1 score respectively.}
\label{Table:persona_eval}
\vspace{-0.2cm}
\end{table}

\begin{table}[t!]
\centering
\small
\begin{tabular}{lccc}
\hline
Simulator Model       & Base Method & SimLearner \\
\hline
Gemma-2-9B-IT     &  0.839  & 0.907   \\
Qwen3-8B  &  0.844  & 0.895   \\
GPT-4o-1120   &  0.851  & 0.930   \\
GPT-4.1-0414 &  0.863  & 0.915   \\
\hline
\end{tabular}
\caption{Concept alignment accuracy between teacher’s guidance and learning outcomes in simulated dialogues.}
\label{Table:tutoring_alignment}
\vspace{-0.2cm}
\end{table}

\subsection{Concept-level Developmental Progress}
To examine concept-level learning trajectories in detail, we compared core ideas learning progress at each grade level between students with and without the SimLearner framework. Figure \ref{Figure_Heatmap} illustrates the developmental progression of NGSS core ideas across grades 1-5. We observe several misalignment issues in Figure 7A, which indicates inappropriate developmental patterns where students access advanced concepts from higher grade levels. In contrast, Figure 7B demonstrates that students with SimLearner exhibit systematic, grade-appropriate concept mastery that follows the intended curricular sequence. Therefore, the SimLearner framework ensures students develop conceptual understanding with core ideas in a structured manner that aligns with curriculum standards.

\subsection{Consistency and Influence of Individual Profiling}

\subsubsection{Personality Traits}
Table \ref{Table:persona_eval} presents the evaluation results on personality consistency. The experiment on four different models demonstrates high consistency in personality traits simulation. Both GPT models and open-source models show stable personality simulation capabilities.
Figure \ref{Figure_Personality_Influence} shows how learning trajectories varied based on three students' personality profiles. With the SimLearner Framework, P1 student demonstrates significantly higher concept understanding than P2 student across nearly all learning stages. These findings align with previous research indicating that Openness, Extraversion, and Agreeableness are associated with better memory performance, while higher Conscientiousness and Neuroticism scores correlate with poorer performance \cite{WARIS201826}.

\subsubsection{Metacognitive Skills}
Given that we also constructed metacognitive skills within the SimLearner framework, we observe how personality traits influence students' metacognitive skills and learning processes. \textbf{P2 student} with high conscientiousness exhibits high level of planning, often responds cautiously like ``\textit{I think I know this, but maybe I should look at it again.}'' \textbf{P1} and \textbf{P3 students} show different behaviors in monitoring. For example, when asked about plant growth, \textbf{P1 student} with high extraversion and low neuroticism often answers questions with confidence, ``\textit{Oh, I totally know this! Plants need water and sun!}'' while \textbf{P3 student} with high extraversion and high neuroticism, shows talkativeness but often with uncertainty ``\textit{Wait, actually... maybe I don't know the answer about this.}'' In contrast, \textbf{P2 student} tend to be more cautious, ``\textit{Um, I think plants need water... but there might be other things too,}'' which shows more conservative but self-reflection on own knowledge. These personality-driven differences in metacognitive processing highlight the role of SimLearner framework in simulating different metacognitive styles.

\subsubsection{Tutoring Agent Alignment}
We also assess the alignment between the teacher's guidance and student learning outcomes in simulated dialogues. As shown in Table \ref{Table:tutoring_alignment}, this measures how accurately the concepts planned by the teacher match the concepts learned by the student across different grade levels. High accuracy indicates that the teacher effectively covers targeted curriculum items and that the simulated responses seldom cause topic drift from the intended tutoring content. The tested models performed well in maintaining the conversation flow and concentrating on the guided science concepts.

\section{Conclusion}
In this work, we present a novel framework for student simulation that addresses key limitations in current LLM-based approaches and educational AI evaluation. Our approach incorporates developmental trajectories through a hierarchical memory mechanism with structured knowledge representation and multi-dimensional learner profiling, enabling more natural simulation of how students gradually build understanding over time and capture the inherent imperfections in human learning processes.
Our experimental results in elementary science learning demonstrate that SimLearner can present curriculum-aligned learning patterns, where the simulated students show grade-appropriate knowledge understanding and maintain developmental progress. Moreover, the integration of personality traits and metacognitive skills creates comprehensive learner profiles that reflect individual differences.
This work paves a new way for building scalable and accessible solutions to evaluate educational AI systems without the constraints of recruiting real participants. The framework can also be extended to various human-AI interaction and educational contexts.

\bibliography{aaai2026}

\clearpage

\appendix

\begin{table*}[t!]
\centering
\begin{tabular}{p{0.9\textwidth}}
\hline
\\
Analyze the student's personality from this tutoring dialogue, and label each Big Five trait as "high" or "low" based on the conversation.\\
\\
Personality trait definitions:
\\
OPENNESS:\\
- High: Creativity in answers; Open to new ideas from teacher; Curiosity and interest in learning
- Low: Lack of creativity in answers; Reluctant to change original ideas; Little interest in learning
\\
CONSCIENTIOUSNESS:\\
- High: Well-organized and logical thinking; Positive attitude toward learning
- Low: Struggling to organize answers; Disengaged in learning; Easily distracted
\\
EXTRAVERSION:\\
- High: Active in conversation; Talkative and enjoyable; Willing to communicate
- Low: Reluctant to talk; Sometimes answering with fillers; Hesitating in answers
\\
AGREEABLENESS:\\
- High: Showing great interest; Empathy and concern for people; Being polite and kind
- Low: Showing little interest in conversation; Not caring about others; Impolite and uncooperative
\\
NEUROTICISM:\\
- High: Feeling anxious; Nervous in conversation; Dramatic shifts in mood
- Low: Emotional stability; Rarely feeling sad or depressed; Confident in answers
\\
Dialogue Sample:\\
\{one\_dialogue\_content\}
\\
\\
Respond with ONLY a JSON object in this exact format:\\
\{\\
"openness": "high" or "low",\\
"conscientiousness": "high" or "low",\\
"extraversion": "high" or "low",\\
"agreeableness": "high" or "low",\\
"neuroticism": "high" or "low",\\
\} \\
\\
\hline
\end{tabular}
\caption{Instruction Example for Big Five Personality Trait Analysis}
\end{table*}

\begin{table*}[t!]
\centering
\begin{tabular}{p{0.9\textwidth}}
\hline
\\
Analyze this learning dialogue and create a episodic memory summary: \\
\\
Dialogue: \\
\{one\_dialogue\_content\} \\
\\
Provide: \\
1. A brief natural language summary (1-2 sentences) \\
2. Key learning insights or confusion points \\
3. Emotional score in range of 0.0 (Negative) to 1.0 (Positive) \\
4. Label the dialogue to the NGSS core idea types. If more than one type, return them all. Directly return the type numbers in the form of "Type XXX". \\
5. Label the mastery level from 1 (no knowledge) to 10 (fully understand) demonstrated by the student of the recognized core idea types. Return a dict format: \{\{'Type XXX': 'SCORE'\}\}. \\
\\
NGSS Core Idea Types: \\
\{all\_concept\_list\_with\_idx\} \\
\\
Format as JSON with keys: summary, insights, emotion, concepts, mastery\_of\_concepts.
\\
\\
\hline
\end{tabular}
\caption{Instruction Example for Episodic Memory Consolidation}
\end{table*}

\begin{table*}[t!]
\centering
\begin{tabular}{p{0.9\textwidth}}
\hline
\\
You are an elementary school student at grade level \{grade\_level\}. \\
\\
Personality Traits:\\
\{personality\_context\} \\
\\
Meta-cognitive Learning Patterns: \\
\{skill\_context\} \\
\\
Recent Learning Experiences: \\
\{recent\_episodes\} \\
\\
Current Concept-level Understanding: \\
\{conceptual\_context\} \\
\\
Reply the teacher based on what you have learned and what you don't understand. Provide short responses.\\
\\
\\
\hline
\end{tabular}
\caption{Instruction Example for the Student Agent of the Elementary School Science Learning}
\label{tab:profile_setting}
\vspace{-2.0cm}
\end{table*}

\begin{table*}[t!]
\centering
\begin{tabular}{p{0.9\textwidth}}
\hline
\\
You are an elementary school teacher at grade level \{grade\_level\}. You are teaching this concept: \{current\_concept\} \\
\\
You are using dialogic teaching approach. This involves co-constructing knowledge through dialogue and collaboration, encouraging exchanging ideas with students and expressing thoughts and opinions, building on prior knowledge and understanding.\\
\\
- If the student shows understanding, acknowledge it and recap their knowledge \\
- If they seem confused, ask guiding questions or provide simpler explanations and hints \\
- Keep responses conversational and age-appropriate \\
- Use examples from everyday life when possible \\
- When the student fully understands the concept, summarized what has been discussed in one sentence \\
\\
\hline
\end{tabular}
\caption{Instruction Example for the Tutoring Agent of the Elementary School Science Learning}
\end{table*}

\end{document}